\documentclass[runningheads,a4paper]{llncs}

\usepackage{amssymb,amsmath}
\usepackage{graphicx}
\usepackage{url}   

\newcommand{\keywords}[1]{\par\addvspace\baselineskip
\noindent\keywordname\enspace\ignorespaces#1}

\newcommand{\be}{\begin{eqnarray}}
\newcommand{\ee}{\end{eqnarray}}
\newcommand{\nn}{\nonumber\\}
\newcommand{\la}{\langle}
\newcommand{\ra}{\rangle}

\newcommand{\nin}{\noindent}

\begin{document}

\mainmatter  

\title{Neural Network for Quantum Brain Dynamics: 4D CP$^1$+U(1)  Gauge Theory on Lattice and its Phase Structure}

\author{Shinya Sakane \and Takashi Hiramatsu \and Tetsuo Matsui}

\institute{Department of Physics, Kindai University\\
Higashi-Osaka, Japan 577-8502\\
\url{}}

%
%

\toctitle{Lecture Notes in Computer Science}
\tocauthor{Authors' Instructions}
\titlerunning{Neural Network for Quantum Brain Dynamics}
\maketitle

\begin{abstract}
We consider a system of two-level quantum quasi-spins and gauge bosons put on a 
3+1D lattice. 
As a model of neural network of the brain functions, these spins describe
neurons quantum-mechanically, 
and the gauge bosons describes weights of synaptic connections.
It is a generalization of the Hopfield model to a quantum network 
with dynamical synaptic weights.
At the microscopic level, this system becomes a model of quantum 
brain dynamics proposed by Umezawa et al.,
where spins and gauge field describe water molecules and
photons, respectively. We calculate the phase diagram of this system 
under quantum and thermal fluctuations, and find that
there are three phases; confinement, Coulomb, and Higgs phases.
Each phase is classified according to the ability to learn patterns and recall them.
By comparing the phase diagram with that of classical networks, we discuss 
the effect of quantum fluctuations and thermal fluctuations (noises in signal propagations) 
on the brain functions.

\end{abstract}

\section{Introduction}
Various functions of the human brain such as awareness, learning, and recalling patterns
have been subjects of intense studies in wide area of science including 
neuroscience, medical science, psychology. A widely adopted approach in these studies is to
model the brain by a neural network (network of neurons) and simulate its static and 
dynamical properties.
A well known example of such network is the Hopfield model \cite{hopfield},
which offers us an interesting 
mechanism of associative memory (recalling memorized patterns of neurons). 

In the Hopfield model, each neuron may have two states (fired or not)
and the state of the $i$-th neuron is described by the Ising (Z(2)) variable $S_i (=\pm1)$ 
($i=1,\cdots, N$). $S_i$ represents the scaled membrane potential as $S_i=1$ (fired) and $S_i=-1$ (not fired). 
The information of memorized patterns of ${S_i}$ is stored here in the 
parameters $J_{ij}$, which are called synaptic weights, through the Hebb's rule \cite{hebb}.
The time development of $S_i(t)\ (t=0,1,2,\cdots$ is a discrete time)
is intrinsically deterministic, but, due to noises in signal propagation, it becomes 
random. This situation is modeled by introducing the energy $E({S_i(t)},{J_{ij}})$ and 
considering statistical mechanics with Boltzmann distribution $P({S_i}) \propto
\exp[-\beta E({S_i},J_{ij})]$ where the effective ``temperature" $T \equiv 1/\beta$ 
starts from zero (no noise) and rises as noise increases. 
Then the system is regarded as an Ising spin system
with random (lomg-range) interactions $J_{ij}$. 
The phase diagram is calculable and consists of three phases;  
spin-ordered phase, spin-disordered phase, spin-glass phase according to 
the values of $J_{ij}$ and $\beta$.  The spin-ordered (ferromagnetic) phase
corresponds to the state of successful recalling of learned patterns of ${S_i}$, 
while the spin-disorders (paramagnetic) phase to failed recalling, and 
the spin-glass phase to failed learning due to more patterns than the capacity.

As the next step, by regarding synaptic weights connecting 
neurons as plastic dynamical variables, various models 
of learning patterns have been proposed \cite{learning}.
In Refs. \cite{z2nn,ijcnn} a set of new networks for learning have been proposed
by promoting synaptic-weight parameters $J_{ij}$ appeared in the Hopfield model
to a dynamical gauge field $J_{ij}(t)$ ($t$ is the time).
The energy of these gauge neural networks respects gauge symmetry. 
Introduction of gauge theory as a model of brain functions
is motivated from the function of synaptic weight itself.
Let us consider the electric signal which starts from the neuron $j$ and arrives at 
the neuron $i$. The electric potential transported by this signal is modulated from 
the initial value $S_j$ at $j$ to $J_{ij}S_j$ at $j$ through the synapse.
The synaptic weight $J_{ij}$ is just the conversion factor of propagating potential.
That is,  $J_{ij}$ expresses relative difference of two frames of potential at $j$ and $i$.
Any quantity having this nature, i.e., a measure of relative orientations of 
local frames, is to be called a gauge field. The gauge symmetry just implies that
observable quantities such as energy should be independent of change of local frames
as it should be. By treating these gauge models as models in statistical mechanics,
we calculated their phase diagrams. Generally, they consist 
of three phases;
confinement phase, Coulomb phase and Higgs phase. Each phase is characterized 
by the ability of learning patterns and recalling them (See Table I). 

Our common sense tells us that the brain functions have
nothing to do with quantum theory (or quantum effect is negligibly small). 
However, as long as our brain is made of atoms 
and molecules at the microscopic level, the microscopic model of the brain 
itself should be 
described in terms of these atoms and molecules.  If we are involved in the enterprise of describing and understanding\\
 
\nin
{\small Table I: Three phases of gauge neural network and abilities of learning and recalling
patterns of $S_i$ \cite{z2nn,ijcnn}. $\la O \ra$ is the Boltzmann average of $O$.
$\la J_{ij}\ra \neq 0$ implies that $J_{ij}$ has small fluctuations around 
the average (given by Hebb's law \cite{hebb}), and the enough 
information of memorized patterns are stored in $J_{ij}$, while $\la J_{ij}\ra = 0$ implies that strong fluctuations wash out such information. Similarly $\la S_i \ra \neq 0$ implies that
$S_i$ sustains an almost definite pattern, while $\la S_i\ra =0$  implies $S_i$ is 
totally unfocused.

\begin{center}
\begin{tabular}{|c|c|c|c|c|} 
\hline
 Phase  &$ \la J_{ij}\ra $ & $ \la  S_i \ra $ &ability of learning &ability of recalling \\ \hline
 Higgs &            $ \neq 0$ &          $\neq 0$         &yes & yes \\ \hline
 Coulomb &       $ \neq 0$ &  $ 0 $ & yes& no  \\ \hline
 Confinement &$ 0 $&$ 0 $& no&no\\ 
\hline
\end{tabular}
\end{center}
}

\nin
 the brain functions by a
framework of physics, our task should be relating such microscopic quantum model to 
widely studied neural networks at the macroscopic  level and calculating the quantum 
effect upon them quantitatively.
This paper concerns these two points.

In Sec. 2 we briefly explain quantum field theory proposed by 
Umezawa et al. \cite{umezawaandtakahashi}
as a model of brain dynamics at the microscopic level. 
It consists of two-level quasi-spin variables describing dielectric dipoles of water molecules and
bosons describing photons inside the brain which mediate the electromagnetic (EM) forces
between dipoles.
We respect the U(1) gauge invariance of EM interaction and introduce 
the CP$^1$+U(1) lattice gauge theory put on a 4D lattice (3 spatial directions and 1 
imaginary-time direction for path-integral quantization)
as its lattice version. Introduction of a lattice model is to discuss an effective model 
at semi-macroscopic scales through renormalization.  
We then discuss that this lattice gauge theory itself 
may be regarded also as an effective GNN after parameters of the model 
are renormalized through coarse graining.

In Sec.3 we calculate the phase diagram of this 4D CP$^1$+U(1) lattice gauge theory
for general parameters and characterize each phase of Table I by measuring electric field, magnetic field, and magnetic monopole density. By considering this model as a GNN, we 
discuss the ability of learning and recalling patterns in each phase,   
and the quantum and thermal(noise) effects upon that ability by referring to 
the results of classical GNN's.

\section{Quantum Brain Dynamics and the 4D CP$^1$ + U(1) Lattice Gauge Theory}

Umezawa et al. \cite{umezawaandtakahashi} proposed a quantum spin-boson model 
that may describe the brain at the microscopic level, and argued that memories 
may be stored in the ordered ground state and low-energy excitations.
They considered a system of $N$ atoms which interact through exchanging bosons.
The $m$-th ago ($m=1,\cdots, N$) is described by 
$s=1/2$ SU(2) pseudo-spin operators $\vec{S}_{m}
=(S_{m1}, S_{m2}, S_{m3})$, and a boson having a 3D momentum $k$ 
and  energy $E(k)$ is described
by canonical annihilation operator $C_k$ and creation operator $C^\dag_k$. 
Its Hamiltonian $H$ is given by
\be
H &=& \sum_{k}E_k C^\dagger_k C_k + \epsilon \sum_{m}
S_{m3}-f\sum_m ( C_m S_{m+} + {\rm H.c.}),
\label{utmodel}
\ee
where $S_{m+}=S_{m1}+iS_{m2}$ is the spin rising operator 
and $C_m$ is the Fourier transform of $C_k$.
Each term expresses energy of bosons,  level splitting 
of spins by external field, and emission and absorption of bosons
and associated spin flips.   
Jibu and Yasue \cite{jibuandyasue} argued that the quasi-spins and bosons 
in Eq. (\ref{utmodel}) have 
explicit counterparts in the human brain; 
each quasi-spin $\vec{S}_m$ describes 
an electric dipole moment of each molecule of bound water (water molecules stand almost still)
and the bosons $C_k$ describe evanescent photons mediating short-range interaction
among dipoles. 

To pursue this interpretation further and improve a couple of
 points of the model (\ref{utmodel}), we introduce a model with the following properties;
(i) manifest U(1) local gauge invariance of EM interaction;
(ii) self-consistently determined photon energy $E(k)$ (massive or massless); 
(iii) a lattice model with a cut-off scale to make renormalization-group
transformation straightforward.
It is a CP$^1$+U(1) lattice gauge theory defined on the 4D hyper-cubic lattice,
a variation of Wilson's lattice gauge theory\cite{wilson} by replacing 
fermonic quark variables by the CP$^1$ spin variables. 
We shall work in the path-integral representation of
the partition function, $Z = {\rm Tr} \exp(-\beta H) $.
The imaginary time $\tau (\in [0,\beta])$ is also
discretized with the lattice spacing $a_0$. 
We use $x = (x_0,x_1,x_2,x_3)$ 
as the site index of the 
4D hypercubic lattice, and  $x_1,x_2,x_3 = 0,1,\cdots, N-1$
and $x_0 =0,1,\cdots, N_0-1$ and $\beta = N_0 \times a_0$.
We use  $\mu=0,1,2,3$ 
as the direction index and also as the unit vector in 
the $\mu$-th direction. 
The lattice spacing $a_\mu=(a_0,a,a,a)$ is regarded as
a parameter to set the scale of the model in the sense of 
renormalization group. 
The $s=1/2$ spins are described by the so-called CP$^1$ (complex projective) 
variables $z_{x\sigma} (\sigma=1,2)$ on each site $x$, 
a two-component complex variables 
satisfying $|z_{x1}|^2+|z_{x2}|^2=1$.
On each link $(x,x+\mu)$ (straight path between two nearest-neighbor (NN) sites), 
we have a U(1) gauge variable,
$U_{x\mu}=\exp(i\theta_{x\mu})$ [$\theta_{x\mu}\in 
(-\pi,+\pi)$]. In the naive continuum limit ($a_\mu \rightarrow 0$), 
it is expressed as $U_{x\mu} = \exp(ig a A_{\mu}(x))$ 
where $A_{\mu}(x)$ is the vector potential and $g$ is the 
gauge coupling constant\cite{wilson}. 
$U_{x\mu}$ measures the relative orientation of the two internal coordinates
which measure the wave function of charged particles at $x$ and $x+\mu$ \cite{wilson}.
Then $Z$ is written as
\be
Z &=& \int[dU][dz]\exp(A[U,z]),\nn
\left[dU\right] &\equiv& \prod_{x,\mu}dU_{x\mu}=\prod_{x,\mu}
\frac{d\theta_{x\mu}}{2\pi},\quad
\left[dz\right]
\equiv \prod_{x}dz_{x1}dz_{x2}\delta(|z_{x1}|^2
 +|z_{x2}|^2-1).
\label{Z}
\ee
 $A[U,z]$ is the action of the model given by
\be
\!\!\!\!A&=&\!\frac{c_1}{2}\!\!\sum_{x,\mu,\sigma}\!\Big(\bar{z}_{x+\mu,\sigma}
U_{x\mu} z_{x\sigma} \!+\mbox{c.c.}\Big) +
\frac{c_2}{2}\!\!\sum_{x,\mu<\nu}\Big(\bar{U}_{x\nu}\bar{U}_{x+\nu,\mu}
U_{x+\mu,\nu}U_{x\mu}\!+\mbox{c.c.}\Big),
\label{A}
\ee
where $c_1$ and $c_2$ are real parameters of the model.
These parameters are regarded to characterize each brain,
i.e., each person has his(her) own values of $c_1$ and $c_2$ 
(and the other parameters for (irrelevant) interactions not 
included here).
The action $A$ is invariant under the following U(1) local
($x$-dependent) gauge transformation;
\be
z_{xa}&\rightarrow& z'_{xa}= e^{i\Lambda_x}z_{xa},\quad
U_{x\mu}\rightarrow U'_{x\mu}= e^{i\Lambda_{x+\mu}}
U_{x\mu}e^{-i\Lambda_x}.
\label{gaugesymmetry}
\ee
Here we note that the partition function $Z$ of (\ref{Z})
is a function
of $\beta c_1$ and $\beta c_2$. Below we set $\beta=1$
in the most of expressions for simplicity. The $\beta$-dependence
is easily recovered by replacing $c_{1(2)}\rightarrow \beta
c_{1(2)}$. 
In the continuum limit $a,a_0\to 0$,
the $c_1$-term of (\ref{A}) becomes the kinetic term 
of $z_x$, while the $c_2$-term becomes the electomagnetic
action $\propto \vec{E}\vec{E}+\vec{B}\vec{B}$.

By applying the renormalization-group transformation to the model (\ref{Z}),
one may obtained an effective theory at the lattice spacings $a_\mu'=\lambda a_\mu$. 
The analysis made for the related models of lattice gauge theory
\cite{Rothebook} shows that the relevant interactions at larger distances are 
the $c_1$ and $c_2$ terms and next-NN terms such as $\bar{z}UUUz, \bar{z}UUUUz$, and no qualitatively different terms emerge. Thus we think that
the  model (\ref{Z}) may be worth to study as an approximation of the effective model 
of neural network for the brain. In this viewpoint, the meaning of variables are as follows;
(i) the CP$^1$ variable $z_{x\sigma}$ is the probability amplitude of  quantum neuron state 
$|S_x\ra = z_{x1}|1\ra_x+z_{x2}|2\ra_x$ where $|1\ra$ and $|2\ra$ are two independent states, such as fired or unfired, and (ii) the U(1) variable $U_{x\mu}=\exp(i\theta_{x\mu})$
is the phase part of wave function of the synaptic connection weight between NN pair
$(x,x+\mu)$.  Therefore, by replacing $z_{x\sigma}$ and $U_{x\mu}$ 
by the neuron variable $S_x$ and the synaptic weight variable $J_{x\mu}$ respectively,
the action $A$ of Eq. (\ref{A}) is viewed 
as the action of GNN at macroscopic level; 
\be
\!\!A\!=\!\frac{c_1}{2} \!\sum \bar{S}_{x+\mu}J_{x+\mu,x}S_x\! 
+\! \frac{c_2}{2} \!\sum J_{x,x+\nu}
J_{x+\nu,x+\mu+\nu}J_{x+\mu+\nu,x+\mu}J_{x+\mu,x}+{\rm c.c.}
\label{gnn}
\ee
We note that its first term $c_1 SJS$ corresponds to the Hopfield energy \cite{hopfield} and the second term $c_2 JJJJ$
describes the reverberating current of signals explained in Ref. \cite{hebb},
which runs along a closed loop $(x\to x+\mu \to x+\mu+\nu \to x+\nu \to x)$.

Of course we recognize that the brain itself is far more complicated than this effective model;
e.g., the network is multilayer with column structure and
the synaptic connections are long-range and asymmetric ($J_{ij}$ and $J_{ji}$ are independent)
with various strengths ($J_{ij} \in {\bf R}$). However, these points can be incorporated systematically
into the present model (\ref{Z}) in the framework of quantum gauge theory
as inputs in the stage of model building , 
and we leave them as future problems.   


\section{Phase Structure of the 4D CP$^1$ + U(1) Gauge Theory}

In this section we study the phase structure of the model (\ref{Z})
by Monte Carlo simulation (MCS)  and 
discuss the effect of quantum and thermal fluctuations upon the ability
of learning and recalling patterns.

\subsection{Phase Diagram}
In our MCS, we consider a hypercubic lattice of size $L^4$ with periodic
boundary condition.
This implies the corresponding ``temperature" $T$ tends to zero $T\to 0$ as
the thermodynamic limit $L \to \infty$ is taken \cite{Rothebook}.
We use standard Metropolis algorithm to generate Markov process
and present the results of $L=16$ with typical sweep number for single run as 
50000+10$\times$5000. Errors are estimated as standard deviation of 10 samples
taken in the last half of each run.
To locate the phase transition point, we calculate the internal energy $U$ 
and the specific heat $C$ defined as the thermodynamic averages as
\be
\la O \ra \equiv \frac{1}{Z}\int[dU][dz]O[U,z] e^{A[U,z]},\ \
  U = \frac{1}{L^4}\la -A \ra, \ \
  C = \frac{1}{L^4}\la (A - \la A \ra)^2\ra,
\ee
where $Z$ and $A$ are given in Eqs. (\ref{Z}), (\ref{A}).
We measure $U$ and $C$ as a function of $c_1$ for a fixed value  of
$c_2$ (and vice versa).
Location of phase transition point is determined from their behavior
as follows;\\

\begin{figure}[t]
\includegraphics[width=5cm]{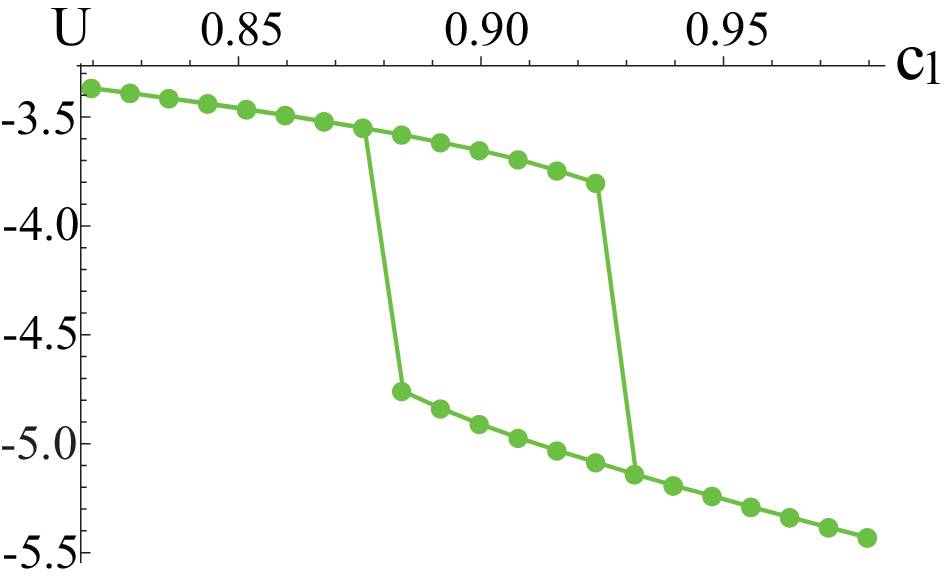}
\hspace{2cm}
\includegraphics[width=5cm]{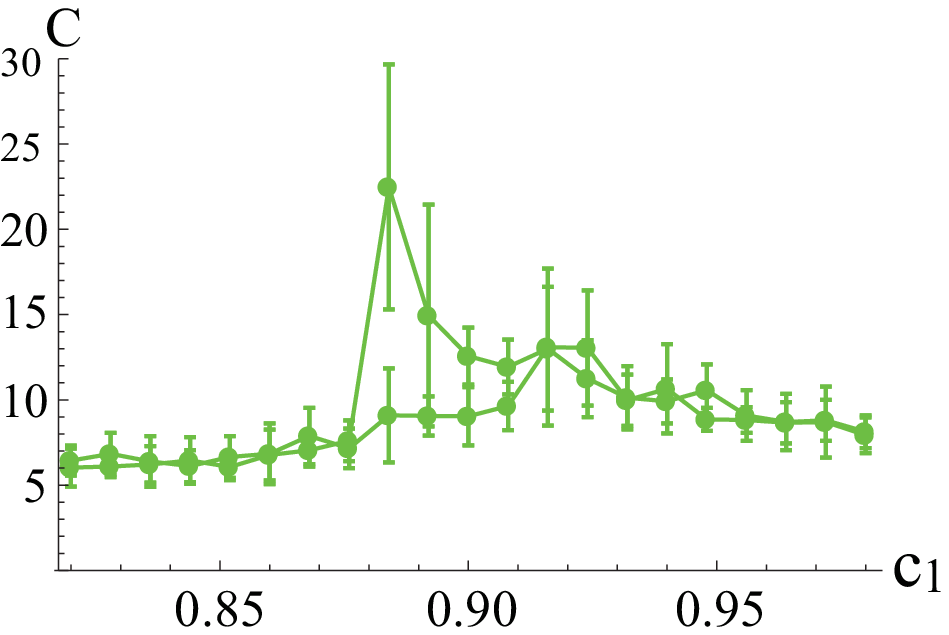}
\caption{$U(c_1)$ (left) and $C(c_1)$ (right) for $c_2=0.9$. $U$ shows a hysteresis
between  $c_1\simeq  0.88 \sim 0.93$. $C$ shows double peaks near the edges
of hysteresis. }
\label{c209}
\end{figure}

\begin{figure}[b]
\includegraphics[width=5cm]{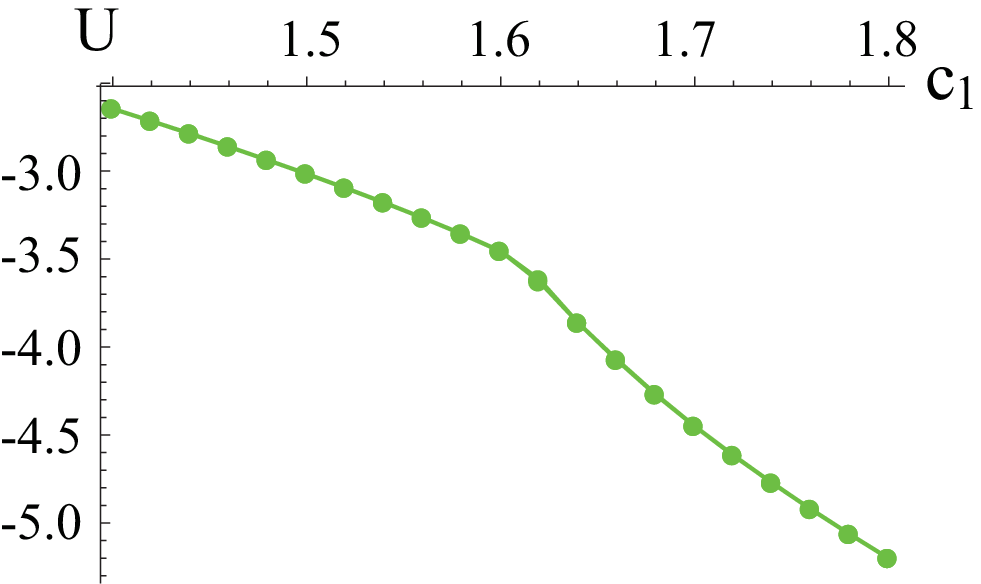}
\hspace{2cm}
\includegraphics[width=5cm]{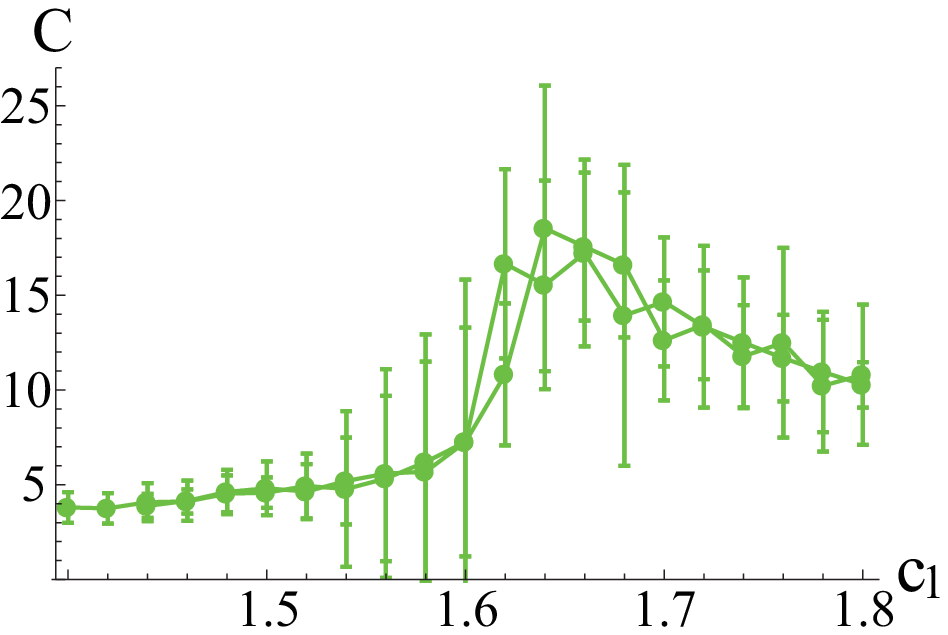}
\caption{$U(c_1)$ (left) and $C(c_1)$ (right) for $c_2=0.4$. $C(c_1)$ shows a peak
at $c_1\simeq 1.64$, at which a second-order transition takes place.}
\label{c204}
\end{figure}

\begin{figure}[t]
\includegraphics[width=5cm]{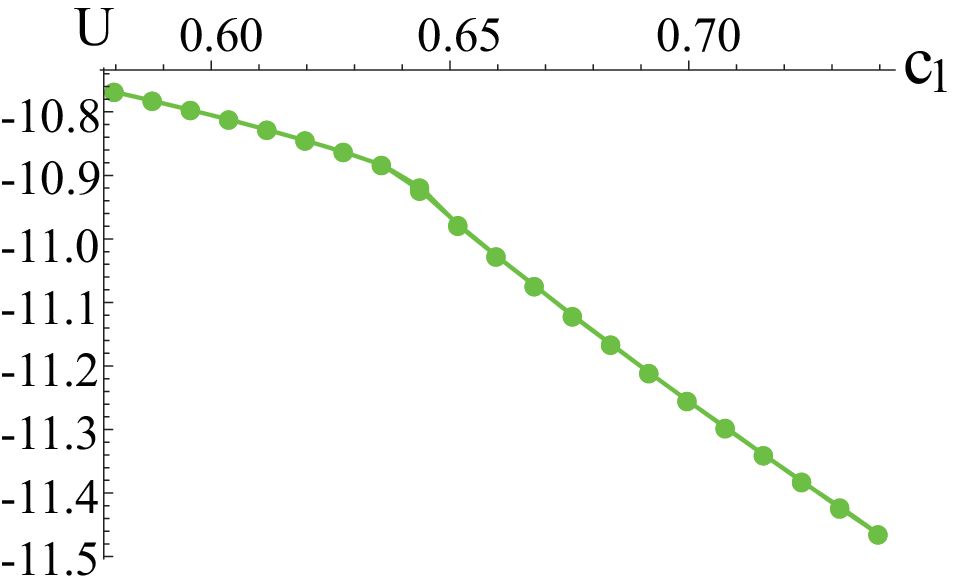}
\hspace{2cm}
\includegraphics[width=5cm]{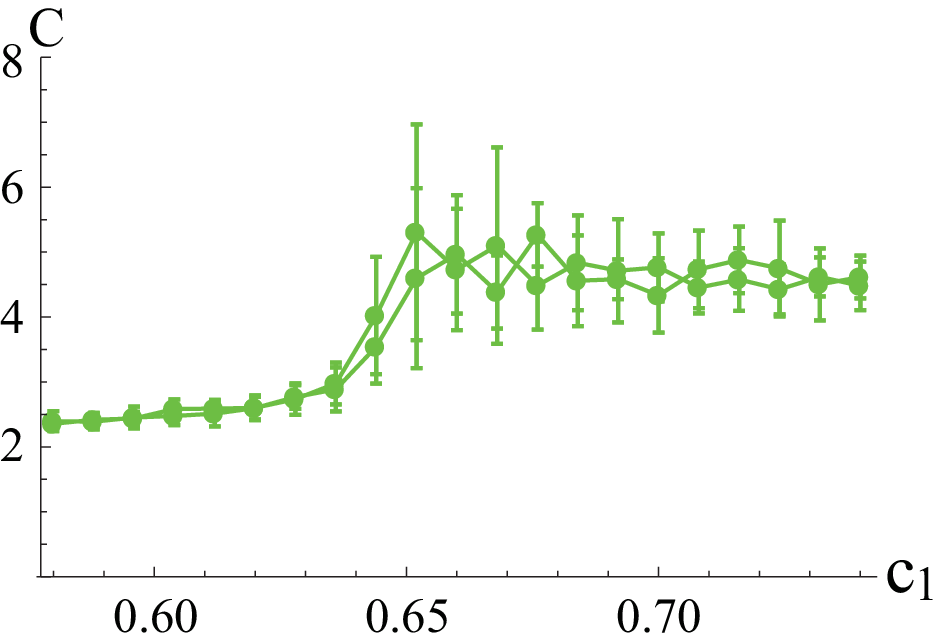}
\caption{$U(c_1)$ (left) and $C(c_1)$ (right) for $c_2=2.0$. $C(c_1)$ has a
 jump at $c_1\simeq 0.65$ which we judge as a gap $\Delta C\neq 0$, 
 implying a second-order transition.}
\label{c220}
\end{figure}

\begin{figure}[h]
\begin{center}
\includegraphics[width= 6cm]{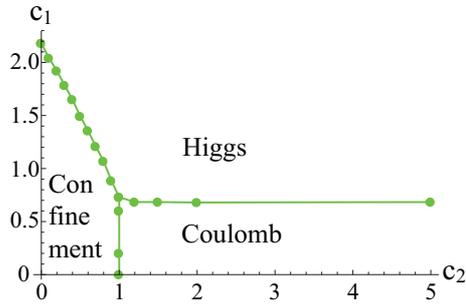}
\caption{Phase diagram of the 4D CP$^1$+U(1) model (\ref{Z}) in the 
$c_2$-$c_1$ plane determined by the MCS for the lattice $L=16$. The transition between Coulomb and Higgs phases is of second-order. The confinement-Coulomb transition is 
of weak first order (almost second order), The confinement-Higgs transition is (i)
first-order near the triple point, i.e., for $0.6 \lesssim c_2 \lesssim 1.0$,
and (ii)  second-order for $c_2 \lesssim 0.6$.  }
\end{center}
\label{PDv1}
\end{figure}

\nin
(i) If $U(c_1)$ shows hysteresis while $c_1$ makes a round trip,
it exhibits a first-order transition.
Such hysteresis effect should diminish as MC runs more sweeps and 
leaves a gap $\Delta U(c_1)$ at the transition point $c_1=c_{1c}$
($\Delta U(c_1)\equiv \lim_{\epsilon\to 0_+}[
U(c_1+\epsilon)$ $- U(c_1-\epsilon)]$).   
\\
(ii) If $U(c_1)$ shows no hysteresis, but $C(c_1)$ has a peak developing
as $L$ increases and/or a gap of $\Delta C(c_1)$ appears at $c_1=c_{1c}$,
a second-order transition takes place there.\\
\nin
As typical examples of these transitions, we show the following three figures;
In Fig. \ref{c209} we show $U$ and $C$ for $c_2= 0.9$. $U$ exhibits a hysteresis curve 
around  $c_1\sim 0.9$ and a first-order transition  takes place.  
In Fig. \ref{c204} we show $U$ and $C$ for $c_2= 0.4$. $C$ exhibits a peak
around $c_1\simeq 1.63$ at which a second-order transition takes place.
In Fig. \ref{c220} we show $U$ and $C$ for $c_2= 2.0$. $C$ exhibits a small jump 
which we take as a sign of a gap $\Delta C$ implying a second-order transition. 

\begin{figure}[t]
\includegraphics[width=4cm]{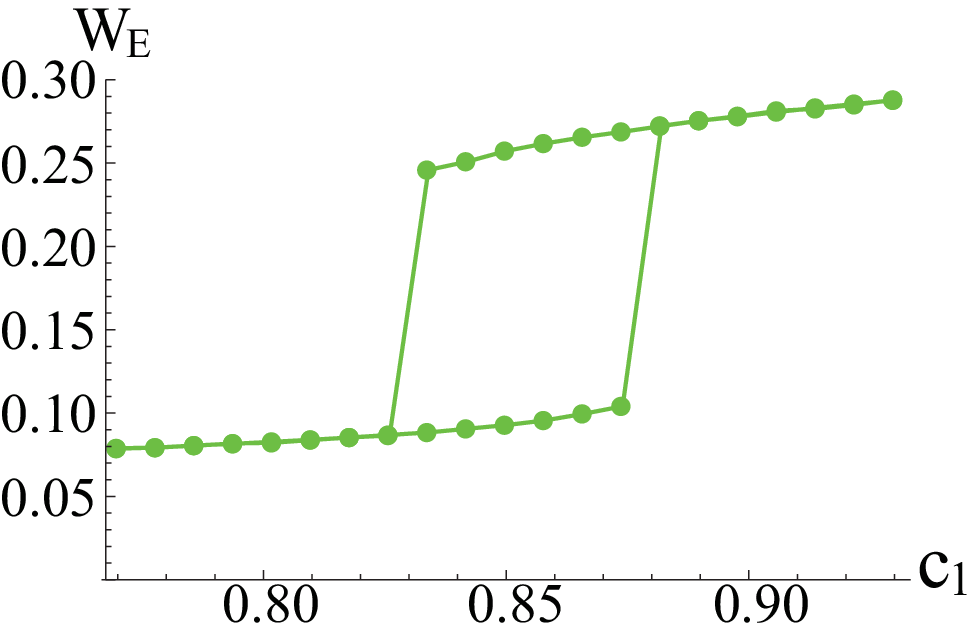}
\includegraphics[width=4cm]{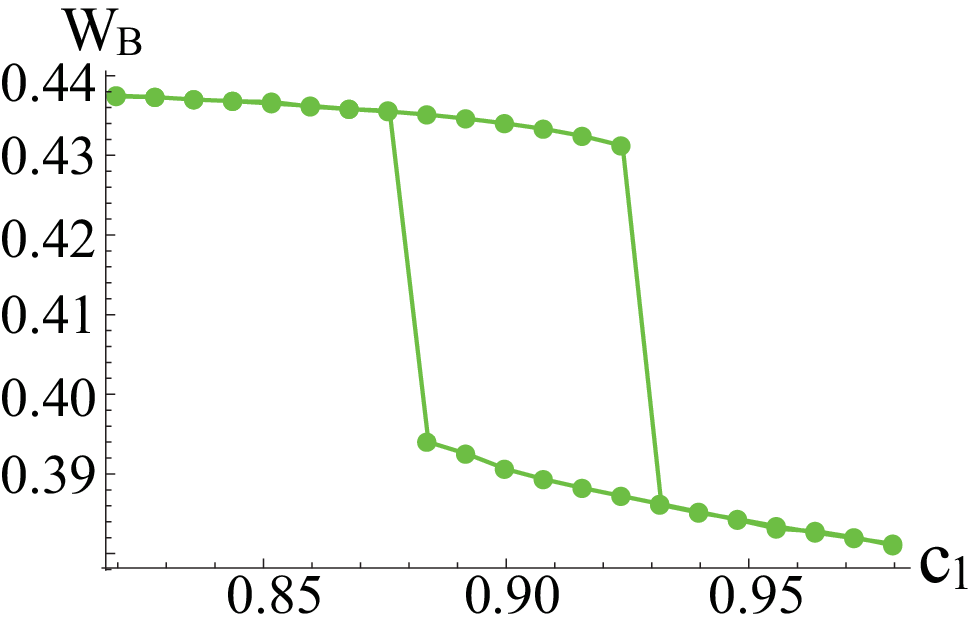}
\includegraphics[width=4cm]{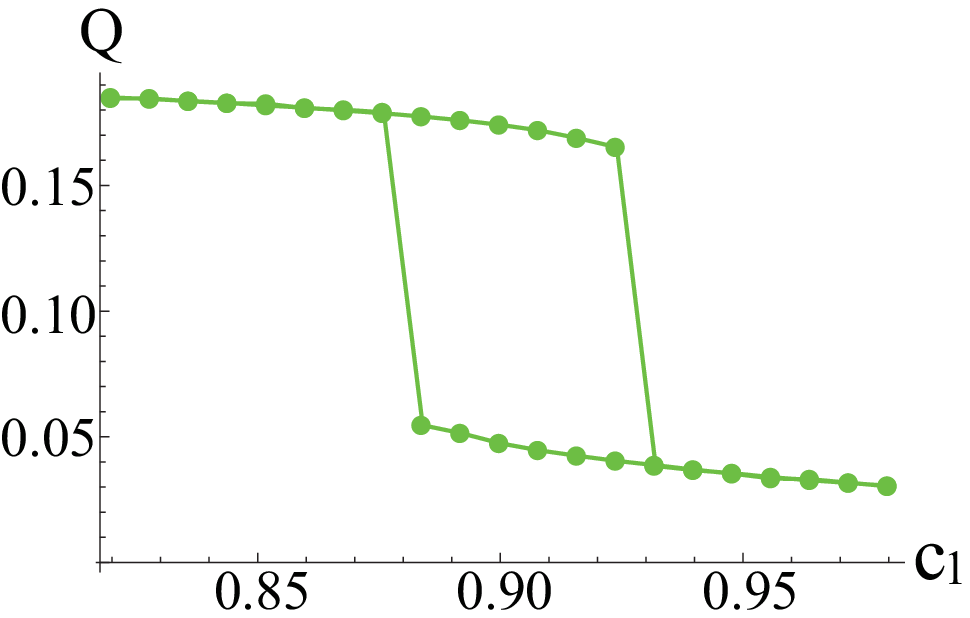}
\caption{$W_E(c_1)$ (left), $W_B(c_1)$ (middle), $Q(c_1)$ (right) for $c_2=0.9$. 
}
\label{c209-2}
\end{figure}

\begin{figure}[t]
\includegraphics[width=4cm]{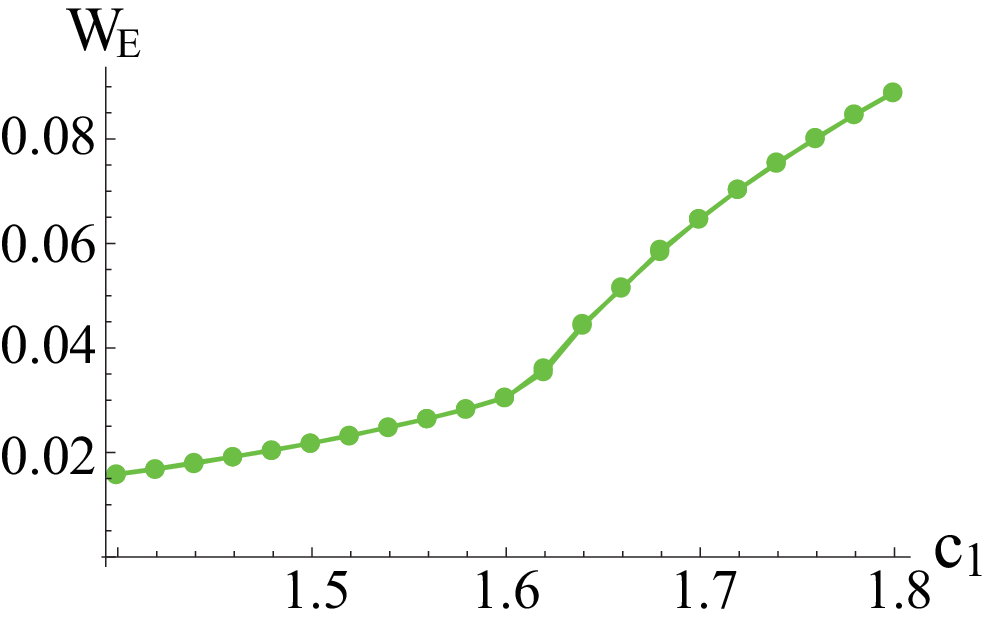}
\includegraphics[width=4cm]{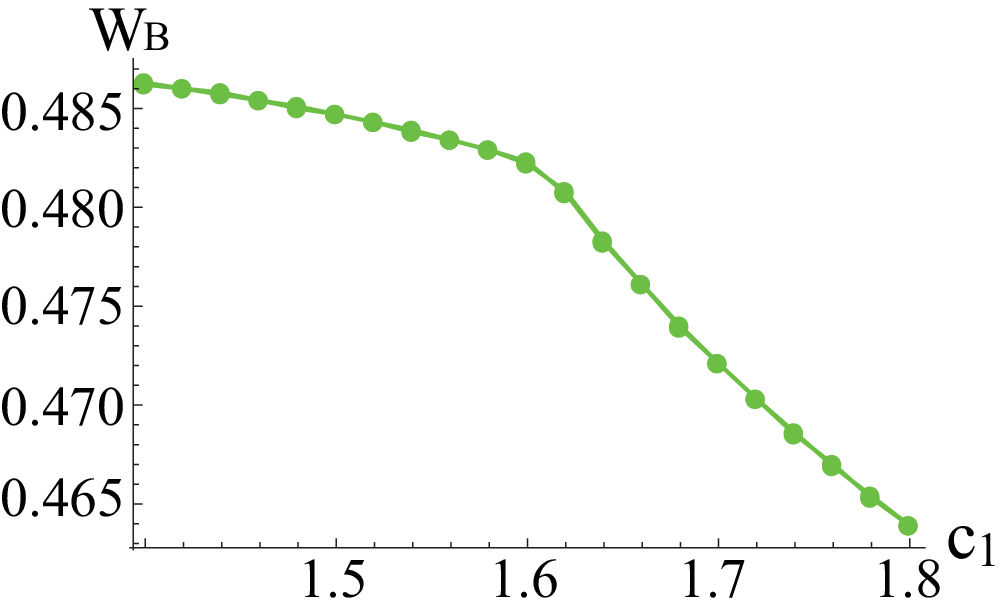}
\includegraphics[width=4cm]{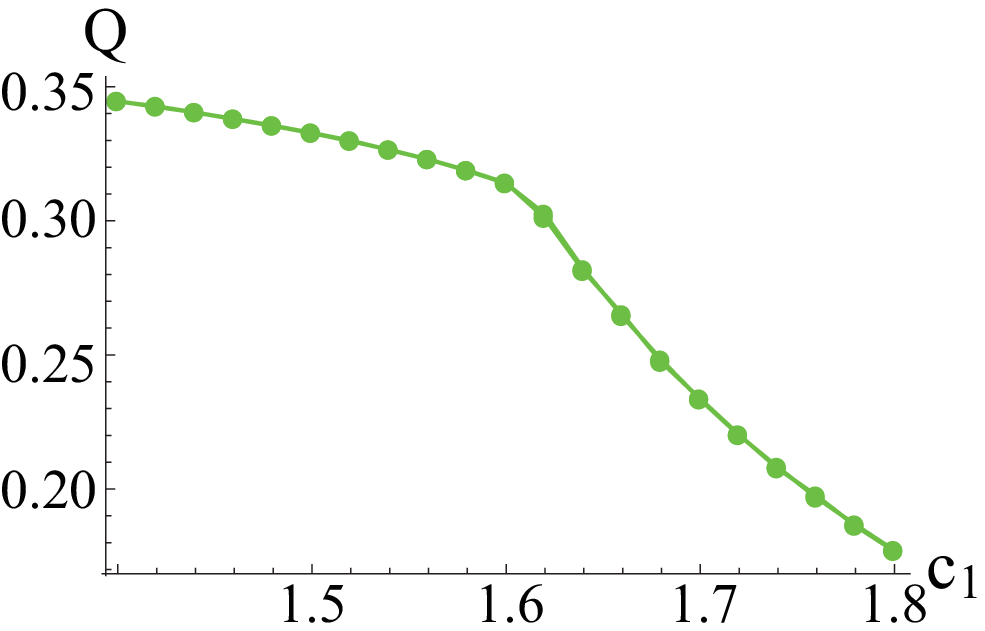}
\caption{$W_E(c_1)$ (left), $W_B(c_1)$ (middle), $Q(c_1)$ (right) for $c_2=0.4$. 
}
\label{c204-2}
\end{figure}

\begin{figure}[b]
\includegraphics[width=4cm]{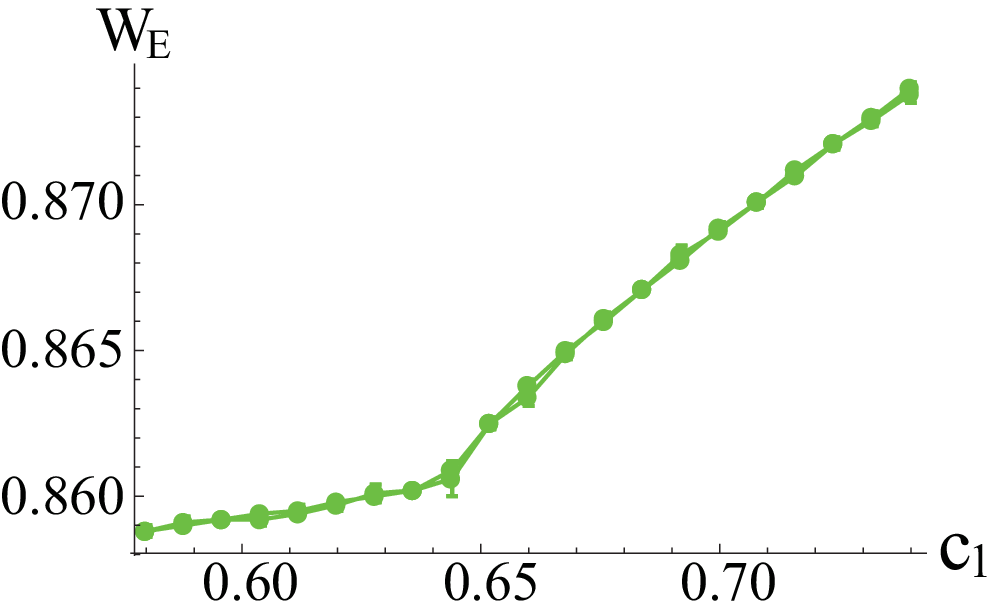}
\includegraphics[width=4cm]{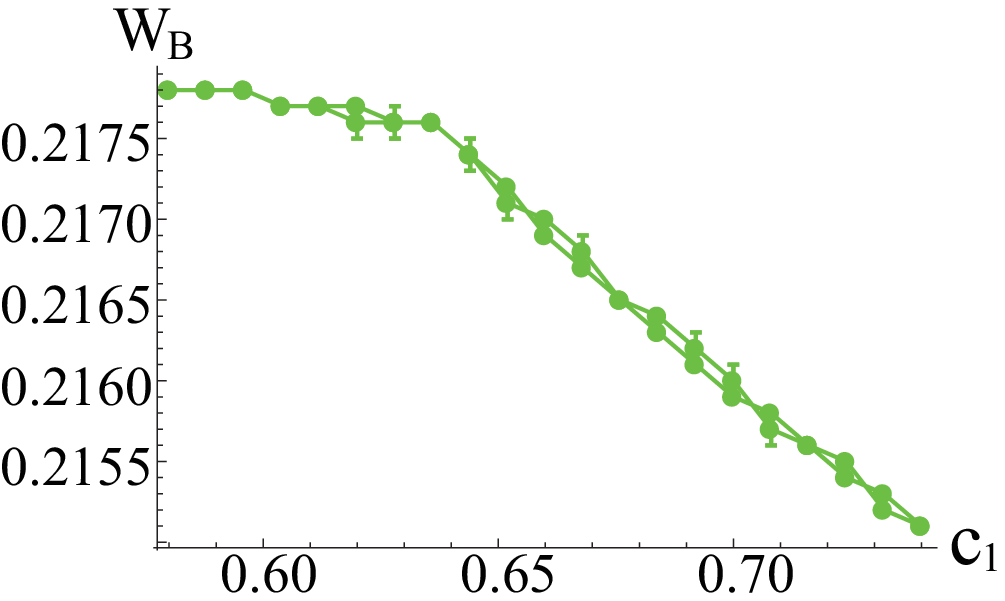}
\includegraphics[width=4cm]{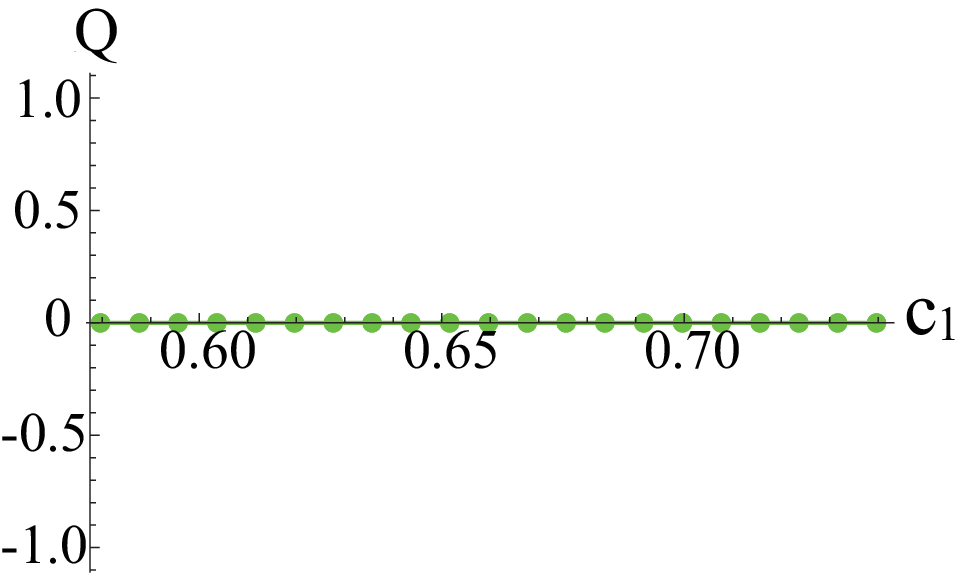}
\caption{$W_E(c_1)$ (left), $W_B(c_1)$ (middle), $Q(c_1)$ (right) for $c_2=2.0$. 
$Q$ almost vanishes here due to strong  suppression of
monopoles due to large $c_2$, while fluctuations in zero-monopole sector generate 
small but finite $W_B$.  
}
\label{c220-2}
\end{figure}

In Fig. 4 
we show the phase diagram in the $c_2$-$c_1$ plane. There are three phases as indicated.
To identify each phase as shown there we measured squared electric field $W_E$,
squared magnetic field $W_B$, and the magnetic monopole density $Q$ \cite{mddef} 
defined as follows;
\be
W_E&\equiv&{1 \over 3L^4}\sum_{x,i}\langle(E_{x,i}-\langle E_{x,i}\rangle)^2\rangle
= \frac{1}{3}\sum_{x, i}
\biggl[c_{2}\langle\cos\theta_{x, 0i}\rangle -c^{2}_{2}\langle\sin^{2}\theta_{x,0i}\rangle\biggr],
\label{E}\nn
W_B&\equiv&\frac{1}{3}\sum_{x, i<j}
\langle\sin^{2}\theta_{x,ij}\rangle,
\label{B}\nn
Q&\equiv&-{1 \over 2}\sum_{i,j,k}\epsilon_{ijk}\la n_{x+i,j,k}-n_{x,jk} \ra 
={1 \over 4\pi}\sum_{i,j,k}\epsilon_{ijk}\la \tilde{\theta}_{x,jk}
-\tilde{\theta}_{x,jk}\ra,
\label{Qx}
\ee
where $i,j,k$ takes $1,2,3$ and we decompose $\theta_{x,ij}=\nabla_i\theta_{xj}-\nabla_j\theta_{xi}$ as
$\theta_{x,ij}=2\pi n_{x,ij}+\tilde{\theta}_{x,ij}, \;\; (-\pi <\tilde{\theta}_{r,ij}<\pi)$.
$n_{x,ij} \in {\bf Z}$ describes nothing but the Dirac string (quantized magnetic flux).
In short, $W_E$ measures the magnitude of fluctuations of electric field $\vec{E}$,
and $W_B$ and $Q$ measure fluctuations of magnetic field
 $\vec{B}={\rm rot}\vec{A}$. Because 
vector potential $\vec{A}$ and $E$ are canonically conjugate each other,
uncertainty principle $\Delta \vec{A} \Delta \vec{E} \sim \Delta \vec{B} 
\Delta \vec{E} \gtrsim$ const. holds. In confinement phase,  $\Delta \vec{E} \simeq$ 0 and 
$\Delta \vec{B}$ is large. In the deconfinement phase such as Coulomb and Higgs phases, 
$\Delta \vec{E}$ is large and $\Delta \vec{B}$ is small. 
$\Delta \vec{B}$ is smaller in the Higgs phase than the Coulomb phase.  
We show these quantities for three values of $c_2$ shown in Figs. \ref{c209},
\ref{c204}, \ref{c220}; $c_2=0.9$ in Fig. \ref{c209-2}, 
$c_2=0.4$ in Fig. \ref{c204-2}, $c_2=2.0$ in Fig. \ref{c220-2}.  
In general, in the small-$c_1$ phase, $W_B$ is large and $W_E$ small,
and in the large-$c_1$ phase, other way around. 
From these properties, it is straightforward to identify three phases
as shown in Fig. 4.

\subsection {Effect of Quantum and Thermal Fluctuations}
To discuss the effect of quantum fluctuations, we introduce a classical model
corresponding to the present quantum model (\ref{Z}).
It is the 4D Z(2) gauge theory defined by the action of Eq. (\ref{gnn})
with the choice $S_x= \pm 1$ and $J_{x,x+\mu}=J_{x+\mu,x}=\pm1$.
These Z(2) variables are discrete and express thermal fluctuations but
no quantum fluctuations. 
In Fig. 8 we show the phase diagrams of these two models obtained by MCS.
It shows that the region of Higgs phase is smaller in the CP$^1$+U(1) model
than in the Z(2) model. Therefore we conclude that the quantum fluctuations
in the present model generally reduce both abilities of learning patterns and recalling them
(see Table I).

So far we considered the case of no noises ($T=0$).
In contrast with $T=0$, the high-temperature limit $T\to \infty$ implies 
$N_0\to1$ in $\beta=N_0 a_0$; i.e.,  the CP$^1$+U(1) model put on the 3D cubic lattice.
Therefore, the effect of noises in signal propagations is estimated by comparing 
the results of the 4D model and 3D model with the same set of variables and action. 
In Fig. 9 we show the 
phase diagrams of the 3D model obtained by MCS\cite{takashima} together with that of 
the 4D model in Fig. 4.
In the 3D model, the confinement-Coulomb transition becomes a crossover and 
Coulomb phase disappears. Furthermore, the region of Higgs phase is smaller than 
that of the 4D model. 
Therefore we conclude that the thermal fluctuations
in the present model generally reduce both abilities of learning patterns and recalling them.

\section{Conclusion}
We introduced the CP$^1$+U(1) gauge theory on a 4D lattice as a microscopic
model of quantum brain dynamics. It describes a system of molecules of bound water
and photons in the brain and respects U(1) gauge symmetry of the electro-

\begin{figure}[h]
\vspace{-0.5cm}
\begin{minipage}{0.45\hsize}
\includegraphics[width=5.5cm]{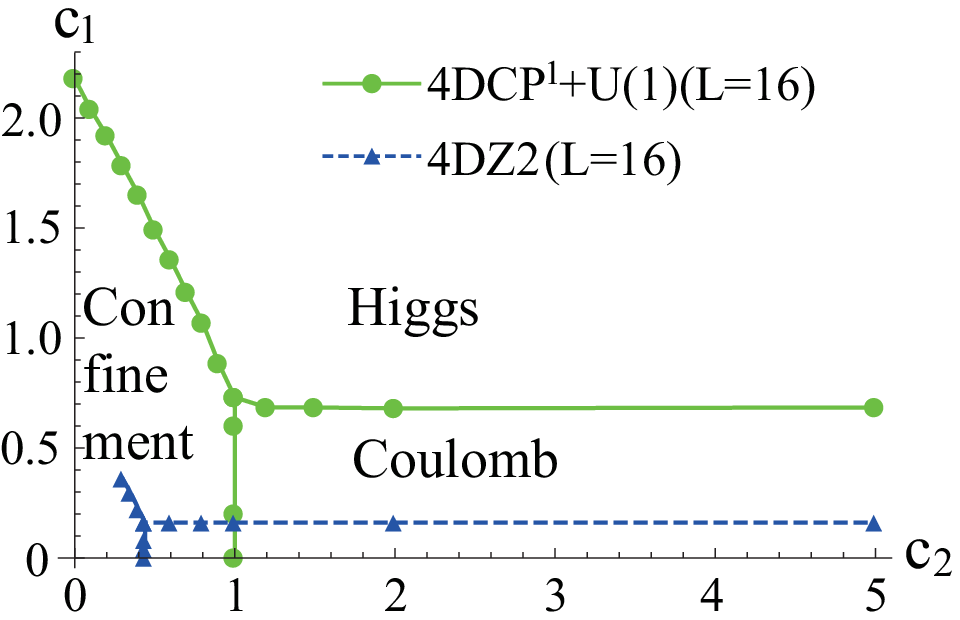} 
\caption{Phase diagrams by MCS for 4D Z(2) model and 4D CP$^1$+U(1) model.
Higgs region is smaller in the CP$^1$+U(1) model.  The transition line 
of the Z(2) model terminates at $c_2\simeq 0.28$.}
\end{minipage}
\hspace{0.1\hsize}
\begin{minipage}{0.45\hsize}
\includegraphics[width= 5.5cm]{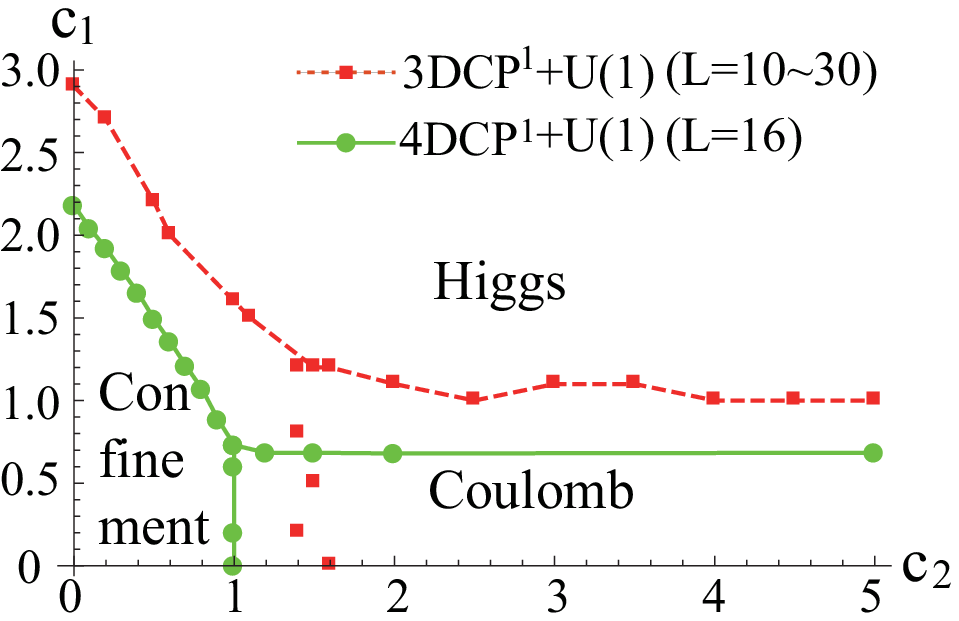}
\caption{Phase diagrams by MCS for 3D and 4D CP$^1$+U(1) lattice gauge models.
Higgs region is smaller in the 3D model. The 3D model has no Coulomb phase and the marks 
at $c_2\simeq 1.4\sim 1.6 $ show the crossover.}
\end{minipage}
\end{figure}

\nin
magnetism.
This model may be regarded also as a neural network of the brain after coarse graining.
We calculated its phase diagram and compared it with related models.
We found that both quantum fluctuations and thermal fluctuations 
by noise reduce the ability of learning and recalling patterns. We plan to
 confirm 
 this point by an explicit simulation of learning processes.   

Finally, we comment on the network structure of the CP$^1$+U(1) model.
To be realistic, the human brain has complicated network structures, such as 
left and right hemispheres, multilayer-structure, column-structure, small-world network,  etc. 
Because the way to coarse-grain the microscopic model is not unique by itself,
additional argument is required to explain the realistic brain structure. 
On this point, it is interesting to define the coarse-grained CP$^1$+U(1) models
on these networks and study their phase diagrams. Although we expect 
the basic three phases appeared in Table I,  the details should be structure-dependent and
shed some light on the study of brain architecture. 

\section*{Acknowledgment}

The authors thank Prof. K. Sakakibara and Dr. Y. Nakano for 
discussion. This work was supported by JSPS KAKENHI Grant
No. 26400412.

\end{document}